\documentclass[aps,prb,twocolumn,showpacs,showkeys,nobibnotes]{revtex4}

\usepackage[dvips]{graphicx}

\newcommand{\degreeC}{$^\circ$C}

\newcommand{\degree}{$^\circ$}

\begin{document}
\title{Ferromagnetism in (In,Mn)As Diluted Magnetic Semiconductor Thin Films Grown by Metalorganic Vapor Phase Epitaxy}%
\date{April 23, 2002}
\author{A. J. Blattner}%
\affiliation{Department of Materials Science \& Engineering \&
Materials Research Center, Northwestern University, Evanston, IL}
\author{B. W. Wessels}
\email{b-wessels@northwestern.edu}%
\affiliation{Department of Materials Science \& Engineering \&
Materials Research Center, Northwestern University, Evanston, IL}

\begin{abstract}
In$_{1-x}$Mn$_x$As diluted magnetic semiconductor (DMS) thin films
have been grown using metalorganic vapor phase epitaxy (MOVPE).
Tricarbonyl(methylcyclopentadienyl)manganese was used as the Mn
source.  Nominally single-phase, epitaxial films were achieved
with Mn content  as high as $x = 0.14$ using growth temperatures
$T_g\geq475$ \degreeC. For lower growth temperatures and higher Mn
concentrations, nanometer scale MnAs precipitates were detected
within the In$_{1-x}$Mn$_{x}$As matrix. Magnetic properties of the
films were investigated using a superconducting quantum
interference device (SQUID) magnetometer. Room-temperature
ferromagnetic order was observed in a sample with  $x=0.1$.
Magnetization measurements indicated a Curie temperature of 333 K
and a room-temperature saturation magnetization of 49 emu/cm$^3$.
The remnant magnetization and the coercive field were small, with
values of 10 emu/cm$^3$ and 400 Oe, respectively. A mechanism for
this high-temperature ferromagnetism is discussed in light of the
recent theory based on the formation of small clusters of a few
magnetic atoms.
\end{abstract}
 \maketitle
\section{Introduction}
An approach to spin-sensitive devices for electronic applications,
that may exhibit improved spin injection, is the use of III-V
diluted magnetic semiconductors (DMSs).\cite{Ohno98,Dietl00} These
alloys, which incorporate a small percentage of magnetic atoms
into the semiconductor host, have been shown to exhibit
ferromagnetic behavior up to 110 K.\cite{Ohno98} However, for
these materials to find widespread applications, the
ferromagnetism should be stable at room temperature. Consequently,
much of the current experimental work is directed towards
increasing their ferromagnetic transition temperature, $T_c$.
Since theory predicts that the transition temperature increases
with the magnetic-ion concentration, efforts have centered on
increasing such concentration.\cite{Ohno00} The solubility of
magnetic ions, however, is often quite low.

III-V DMS have been grown using low-temperature molecular beam
epitaxy (LT-MBE) at temperatures lower than 300 \degreeC\ to
prevent phase separation. Nevertheless, recently we have
demonstrated the growth of single-phase, ferromagnetic
In$_{1-x}$Mn$_x$As, with $x\leq0.14$, using metalorganic vapor
phase epitaxy (MOVPE) at temperatures as high as 520
\degreeC.\cite{Blattner01a} In this report, we present the results
of our investigation of the magnetic properties of these films. We
have observed room-temperature ferromagnetic behavior in a
nominally single-phase $p$-type In$_{0.9}$Mn$_{0.1}$As film.  In
contrast, a comparable high $T_c$ has not been observed in
single-phase (In,Mn)As grown using LT-MBE.  A possible mechanism
for the observed high-temperature ferromagnetism, based on
transition metal clustering at the atomic level, will be
discussed.
\section{Epitaxy of (I\lowercase{n},M\lowercase{n})A\lowercase{s}}
(In,Mn)As films were prepared using atmospheric pressure MOVPE, as
described earlier.\cite{Blattner01a} Films were grown at
temperatures between 475--520 \degreeC\ on semi-insulating
GaAs(001) substrates.  The precursors used were trimethylindium
(TMIn), tricarbonyl(methylcyclopentadienyl)manganese (TCMn) and
0.3\% arsine (AsH$_3$) in hydrogen.  Phase composition was
determined using double-crystal x-ray diffraction (XRD) using Cu
K$\alpha_1$ radiation. Energy dispersive x-ray spectroscopy (EDS)
was used to determine the Mn concentration in the films.

\begin{figure}[b!]
\centering
\includegraphics[width=\columnwidth,keepaspectratio=true,draft=false,clip=true]{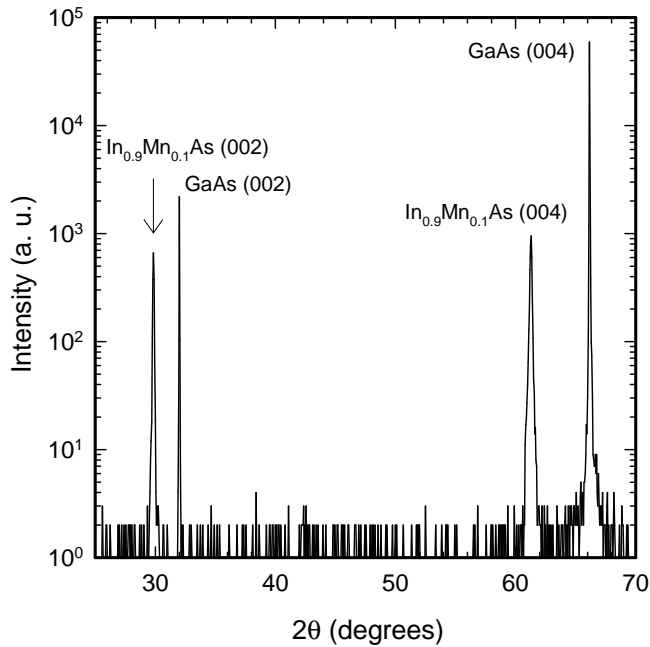}
\caption{\label{fig1}$\theta$-2$\theta$ x-ray diffraction scan of
In$_{0.9}$Mn$_{0.1}$As/GaAs(001) showing the re\-spec\-tive  (002)
and (004) re\-flect\-ions.  No evidence of MnAs second phase was
observed.}
\end{figure}

\begin{figure}
\centering
\includegraphics[width=\columnwidth,keepaspectratio=true,draft=false,clip=true]{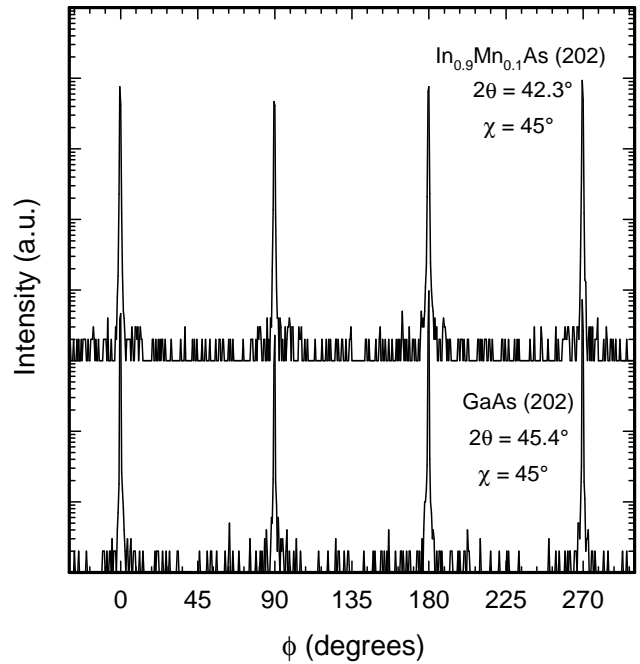}
\caption{\label{fig2}Azimuthal, $\phi$, x-ray diffraction scans
for the \{202\} reflections of In$_{0.9}$Mn$_{0.1}$As and GaAs
indicating the degree of epitaxy with the GaAs substrate.  The
angle of the \{202\} diffracting planes with respect to the sample
surface is $\chi=45$\degree.}
\end{figure}

Figure \ref{fig1} shows the $\theta$-2$\theta$ x-ray diffraction
pattern with the zinc-blende (004) and (002) reflections for a 300
nm thick In$_{0.9}$Mn$_{0.1}$As film grown at 520 \degreeC.
No other peaks were observed, indicating that the alloy was phase
pure to within 0.1 volume percent.  The rocking-curve FWHM for
this film was 0.25\degree. Azimuthal ($\phi$) scans for the
\{202\} reflections of the film and substrate were recorded to
confirm epitaxy. Figure \ref{fig2} shows the four-fold-symmetric
diffraction pattern for the \{202\} reflections of the (In,Mn)As
film compared with that of the GaAs substrate.
Excellent epitaxial alignment is observed despite the $\sim7\%$
lattice mismatch between the film and the substrate.

MnAs was observed in films grown at temperatures lower than 475
\degreeC\ and for Mn concentrations greater than $x=0.14$. Growth
at temperatures higher than $\sim530$ \degreeC\ resulted in very
little deposition presumably due to increased indium and manganese
desorption from the surface.

\section{Magnetic Properties}
\subsection{Magnetic field dependence}
\begin{figure}[b!]

\centering
\includegraphics[width=\columnwidth,keepaspectratio=true,draft=false,clip=true]{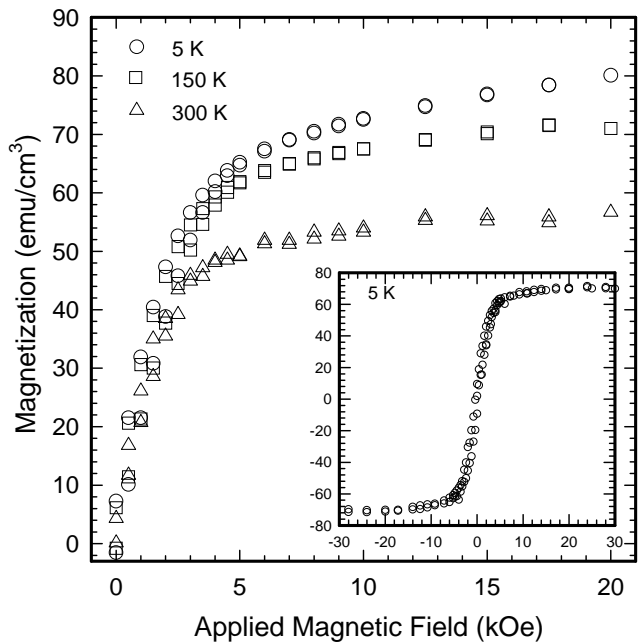}
\caption{\label{fig3}Magnetization as a function of applied magnetic field for
a In$_{0.9}$Mn$_{0.1}$As sample (AJB070) measured at 5, 150, and
300 K. Magnetic field was applied perpendicular to the plane of
the film. Inset is the complete hysteresis loop at 5 K.}
\end{figure}

The magnetic properties of the (In,Mn)As thin films were measured
using a SQUID magnetometer. The magnetic field was applied
perpendicular to the plane of the film along the easy axis of
magnetization for (In,Mn)As.\cite{Munekata93} The substrate
diamagnetic contribution was subtracted from the total
magnetization signal. The remaining data consisted of the
paramagnetic and ferromagnetic contributions from the film.  The
total magnetization is given by:
\begin{equation}
M_{\mathrm{tot}}(T,H)=M_F(T,H)+[\chi_p(T)+\chi_d(GaAs)]H
\end{equation}
where $M_F(T,H)$ is the ferromagnetic component of the film,
$\chi_p$ is the paramagnetic susceptibility of the film, $\chi_d$
is the substrate diamagnetic susceptibility, $T$ is the
temperature, and $H$ is the applied magnetic field.  The magnetic
response of a nominally single-phase In$_{0.9}$Mn$_{0.1}$As film
is shown in Figure \ref{fig3} for applied magnetic fields up to 20
kOe and temperatures of 5, 150 and 300 K.
The inset of Figure \ref{fig3} shows the complete hysteresis loop
for this sample at 5 K. At this temperature the measured
saturation magnetization normalized to the film volume was 62
emu/cm$^3$ with a remanence of 10 emu/cm$^3$ and a coercive field
of 400 Oe. When the temperature was increased to 300 K, the
saturation magnetization decreased only to 49 emu/cm$^3$.

\subsection{Temperature dependence}
\begin{figure}
\centering
\includegraphics[width=\columnwidth,keepaspectratio=true,draft=false,clip=true]{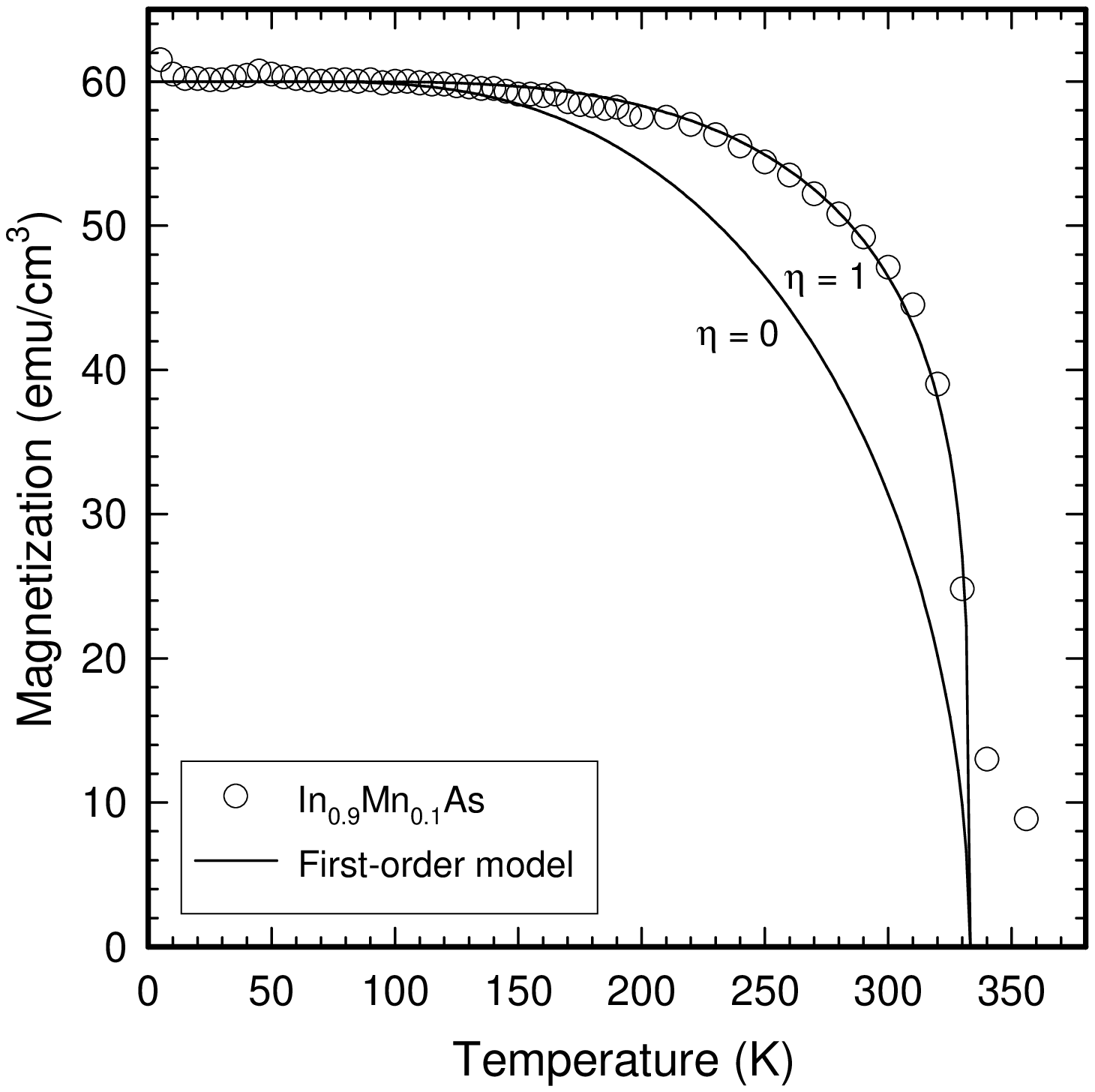}
\caption{\label{fig4}Temperature dependence of the magnetization for the
In$_{0.9}$Mn$_{0.1}$As sample (AJB070) compared with the model of
Bean and Rodbell for $J=1/2$ and $P=0$.\cite{Bean62} Magnetic
field strength was 10 kOe an was applied perpendicular to the
plane of the film. A $T_c=333$ K and $M_s=60$ emu/cm$^3$ were used
for the fitting.}
\end{figure}

The magnetization as a function of temperature was measured for
the single-phase In$_{0.9}$Mn$_{0.1}$As sample discussed above.
The sample was zero-field cooled and then subjected to a 10 kOe
applied magnetic field perpendicular to the plane of the film.
After subtraction of the diamagnetic substrate contribution, the
resulting magnetization is shown in Figure \ref{fig4}.

A Curie temperature of 333 K was measured for this sample.  Such a
high Curie temperature for a III-V dilute magnetic semiconductor
has been previously attributed to the presence of
MnAs\cite{Molnar91} which has a Curie temperature of 318 K.
However, the presence of MnAs was not observed by x-ray
diffraction in this sample as shown in Figure \ref{fig1}. If a
second phase were present, the measured magnetization of 62
emu/cm$^3$ would correspond to a MnAs volume fraction of
approximately 10\%.  This should certainly be observable by x-ray
diffraction.  If on the other hand the MnAs was present as
nanoprecipitates, the magnetization would be considerably smaller
as observed in (Ga,Mn)As films.\cite{Boeck96} Thus the
magnetization data is consistent with the alloy film being a
single-phase.  Currently, we are investigating precipitate
formation further through transmission electron microscopic
analysis.

The value of the saturation magnetization of (In,Mn)As is given by
$M_s=N_\mathrm{Mn}g\mu_BJ_{\mathrm{Mn}}$, where $N_\mathrm{Mn}$ is
the nominal concentration of Mn ions, $g$ is the Lande\'{e} factor
and is equal to 2 for Mn, $\mu_B$ is the Bohr magneton, and
$J_\mathrm{Mn}$ is the spin of Mn.  The measured $M_s$ using
$x=0.1$, corresponding to $N_\mathrm{Mn}$=1.8$\times$10$^{21}$
cm$^{-3}$, gives a value of $\mu=3.6\mu_B$.   This corresponds to
a value of $J_\mathrm{Mn}$ between 4/2 and 3/2. This is
interpreted as an indication of the presence of valence states
other than Mn$^{2+}$.\cite{Ohno91}

The temperature dependence of the magnetization was compared to
the Brillouin function for different values of the total angular
momentum, $J$.  The  lack of agreement between the Brillouin
function and the measured data above 150 K indicates that the
transition to the ferromagnetic state may not be second order, in
which the change in magnetization value is continuous with
increasing temperature. It may be first order.

Magnetic disordering as a first order phase transformation has
previously been reported and a model to describe this behavior was
developed by Bean and Rodbell.\cite{Bean62}  The model predicts
that if the exchange interaction is a strong function of the
inter-atomic spacing and the lattice is compressible, then the
ferromagnetic to paramagnetic transition will be first-order.  By
introducing a strain energy term into the free energy formulation
and minimizing with respect to the reduced magnetization, it was
shown\cite{Bean62} that the temperature dependence of the
magnetization for $J=1/2$ is given by
\begin{equation}\label{eq2}
T/T_o=(\sigma\tanh^{-1}{\sigma})(1+\eta\sigma^2/3-PK\beta)
\end{equation}
where $\eta\equiv 3/2NkKT_c\beta^2$, $\sigma$ is the reduced
magnetization (normalized to the saturation magnetization), $N$ is
the number of magnetic dipoles per unit volume, $k$ is the
Boltzmann constant, $K$ is the compressibility, $\beta$ is the
slope of the dependence of $T_c$ on volume, and $T_o$ is the Curie
temperature if the lattice were not compressible.  $P$ and $T$ are
the pressure and temperature respectively.

For $\eta<1$, the transition is second order, and Equation
\ref{eq2} reduces to the Brillouin function for $\eta=0$.  For
$\eta>1$, the transition is first order and the magnetization
exhibits a discontinuity at the transition temperature.  The
measured magnetization was compared with the model of Bean and
Rodbell for different values of $\eta$ as shown in Figure
\ref{fig4}.
As can be seen, the best fit occurred for $\eta=1$. This indicates
that the phase transformation for the alloy has an intermediate
character between first and second order.

A deviation between theory and experiment at temperatures above
the Curie temperature was also observed. This was addressed in the
initial work by Bean and Rodbell and is a consequence of the
long-range interaction implicit in the model.\cite{Bean62}
Molecular field theory which was used in the model, indicates no
order above the Curie temperature for a normal ferromagnet.
Experimentally, however, significant short-range order can be
observed.

\begin{figure}
\centering
\includegraphics[width=\columnwidth,keepaspectratio=true,draft=false,clip=true]{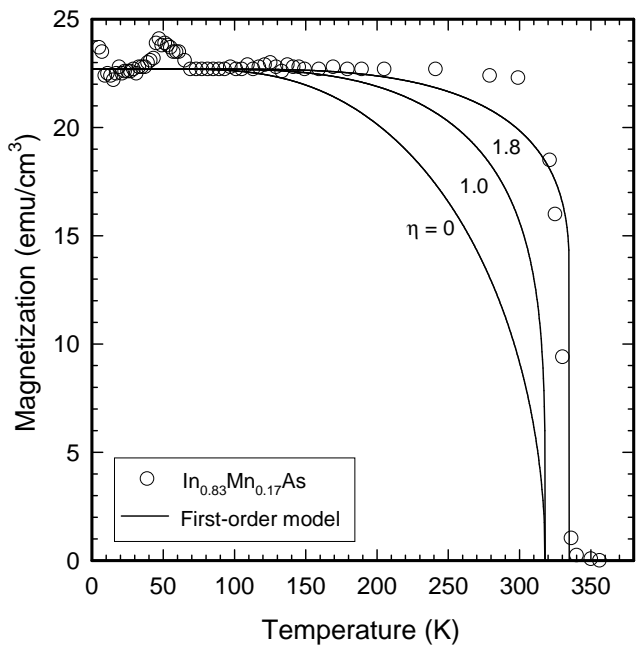}
\caption{\label{fig5}Temperature dependence of the magnetization for a two
phase sample (In$_{0.83}$Mn$_{0.17}$As + MnAs) (AJB076) compared
with the model of Bean and Rodbell for $J=1/2$ and $P=0$. Magnetic
field strength was 1 kOe and was applied perpendicular to the
plane of the film. A $T_c=318$ K and $M_s=23$ emu/cm$^3$ were used
for fitting.}
\end{figure}

For comparison, the ferromagnetic to paramagnetic phase transition
of MnAs has been shown to be
first-order\cite{Bean62,Chernenko99,Pytlik85} with a value of
$\eta=2$.\cite{Bean62} We have measured the temperature dependence
of the magnetization for an (In,Mn)As sample which exhibited MnAs
phase formation. The magnetization is shown in Figure \ref{fig5}
as a function of temperature.
In this case the temperature dependence appears to be clearly
first-order. Fitting of the magnetization data using the
first-order model resulted in a value of $\eta=1.8$. Thus it can
be seen that the temperature dependence of the magnetization for
(In,Mn)As/MnAs two phase alloys differs from that of (In,Mn)As
single-phase alloys.  The temperature dependent magnetization of
In$_{0.9}$Mn$_{0.1}$As sample shown in Figure \ref{fig4} is closer
to second-order and cannot simply be explained by the presence of
MnAs nanoprecipitates.
\subsection{Origin of Ferromagnetism}
The ferromagnetism observed in III-V DMS films has been previously
explained within the framework of sp-d exchange interactions
between the Fermi sea and the localized magnetic
moments.\cite{Dietl01} Similar to RKKY theory, this hole-mediated
ferromagnetism theory predicts that the Curie temperature is
dependent upon both magnetic ion concentration and hole
concentration.  Using this model Dietl et al. predicted that the
transition temperature for (In,Mn)As with a Mn concentration of
5\% should be 35 K, which is much lower than that observed in the
present study. One possible explanation for the difference is that
Mn in the alloy is not randomly distributed on In sites but is
present as atomic clusters. This is supported by the recent
calculations of van Schilfgaarde and Mryasov based on local
density functional theory.\cite{Schilfgaarde01} These calculations
predict that due to the strong attractive coupling between the Mn
ions and the semiconductor cation nuclei, there is a large driving
force for the formation of Mn clusters consisting of two or more
Mn atoms located at nearest-neighbor cation sites.\footnote{Mn
substitutes on the In sublattice forming clusters containing 2 or
3 Mn atoms.  For magnetic ``dimers'', the Mn atoms are located at
nearest-neighbor cation sites as in Ref.
\onlinecite{Schilfgaarde01}.} These atomic clusters in turn can
stabilize the ferromagnetic state; according to this model, the
presence of free carriers is not required. Although the model does
not consider kinetic effects, which may determine whether Mn
clustering occurs, the growth temperature used for our films is
likely sufficient to overcome any kinetic barrier to cluster
formation. Indeed clustering of the transition-metal ion has been
observed in other III-V DMSs grown at high
temperatures\cite{Haneda00,Soo01a,Soo01} and may also occur in
other materials that have been shown to exhibit ferromagnetic
behavior while having $n$-type
conductivity.\cite{Reed01,Overberg01} Presently, we are pursuing
extended x-ray absorption fine structure (EXAFS) measurements of
these films to determine  the local environment of Mn.

\section{Conclusions}
In conclusion, epitaxial, single-phase In$_{0.9}$Mn$_{0.1}$As
films have been grown using metalorganic vapor phase epitaxy at
temperatures as high as 520 \degreeC. Temperature- and
field-dependent magnetization measurements indicated a
single-phase film (as measured by x-ray diffraction) to be
ferromagnetic with a Curie temperature of 333 K. Modelling of the
magnetization data indicated that the
ferromagnetic-to-paramagnetic phase transition was intermediate
between first and second order. The origin of the observed
ferromagnetism was discussed in the light of a new theory based on
the formation of transition-metal clusters.

\section{Acknowledgements}
Extensive use was made of the Materials Research Center facilities
at Northwestern University.  This work was supported by the
National Science Foundation through the MRSEC program under grant
number DMR-0076097.




\end{document}